\documentclass[preprint, prd, aps, floatfix, nofootinbib,
a4paper]{revtex4-1}

\usepackage{graphicx}
\usepackage{amsmath, amssymb, amsfonts, amsthm, latexsym, mathrsfs, bbm, tensor}
\usepackage{bm}
\usepackage{hyperref}
\usepackage[T1]{fontenc}
\usepackage[utf8]{inputenc}
\usepackage{lmodern}

\usepackage{etoolbox}
\apptocmd{\thebibliography}{\raggedright}{}{}

\newcommand{\scri}{\mathscr{I}}
\newcommand{\thorn}{\text{þ}}

\newcommand{\tvect}[2]{\ensuremath{\big[\negthinspace\begin{smallmatrix}#1\\#2\end{smallmatrix}\negthinspace\big]}}
\newcommand\addnumber{\addtocounter{equation}{1}\tag{\theequation}}

\begin{document}

\title{Twistors, charge structure, and BMS symmetries}
\author{Alex Goodenbour}
\email{alex.goodenbour@outlook.com}
\noaffiliation
\date{\today}

\begin{abstract}
  Corresponding to the Bondi-Metzner-Sachs (BMS) symmetry algebra of asymptotically-flat
  spacetimes are a set of \emph{BMS charges}. These are formally
  constructed via the symplectic formalism of Wald and Zoupas, but the
  same charge expression may be arrived at by the simpler twistorial
  procedure of Dray and Streubel. Here, we formalize the connection
  between twistors and asymptotic symmetries which underlies the
  Dray-Streubel charge by demonstrating an \emph{isomorphism} between
  twistors in flat spacetime and twistors on \emph{radiation-free}
  sections of \(\mathscr{I}^+\). In the corresponding formalism, the
  Dray-Streubel charge finds a natural reinterpretation as exactly the
  part of Penrose's twistorial charge which is invariant with respect
  to a certain gauge transformation. Furthermore, we argue that the
  twistorial picture of the radiative phase space, properly
  formalized, provides a tool alongside the symplectic formalism or
  shear structure for analyzing radiative data on \(\mathscr{I}^+\).
\end{abstract}

\maketitle

\section{Introduction}
In a generic curved spacetime, in the absence of a preferred direction such as a
Killing vector field or a fluid 4-velocity, there is no obvious way to define charges
like energy and angular momentum. Charges tend to be defined with respect to
symmetries of the background and outside of certain exact solutions, general
relativity admits no such symmetries. As a consequence, there can be no grounds
for comparison between any two spacetimes without additional structure since
there is no phase space of observables that is shared by all spacetimes.

The advancement that would recover the ability to compare spacetimes came out of
the attempt to covariantly describe gravitational radiation beginning with Bondi
and culminating in Penrose's procedure of conformal compactification
\cite{BondiVIII,GravitationalWBondi1973,penrose1965zero}.  The abstract manifold
\(\scri\) emerges as the place at which gravitational radiation can be
unambiguously defined. It forms a universal structure shared by all spacetimes
which admit a smooth conformal boundary. As shown by Ashtekar and Streubel
\cite{SymplecticGeom1981,OnTheSymplectAshtek1982,AUnifiedTreatAshtek1978}, this
universal structure is a symplectic structure and so it can be thought of as a
phase space shared by all \emph{asymptotically-flat} spacetimes.

We would expect that a charge structure should emerge with respect to symmetries
of this universal structure and so it is significant that the symmetry group of
the universal structure is the infinite-dimensional BMS group, corresponding to the
Poincaré group with the translations replaced by the infinite-dimensional
supertranslations \cite{AsymptoticSymmSachs1962}.

Taking the position that the infinite-dimensional enhancement of the symmetry
group in the infrared is something to be taken seriously, a set of adjoint
\emph{BMS charges} may be defined
\cite{ds,LinkagesInGenGeroch1981,waldzoupas,Streubel1978}. Two of these
prescriptions are formally equivalent and have all of the properties that one
would reasonably expect. The most conceptually satisfactory of these is the
Wald-Zoupas prescription which defines a charge in terms of the explicit
symplectic structure at null infinity, taking into account the leaking of
symplectic current through the boundary \cite{waldzoupas,grant2022wald}.

The second is the Dray-Streubel prescription which exploits the spin-lowering
properties of the twistor equation to define a charge in analogy with Maxwell
electromagnetism \cite{Penrose1982wp, ds,Dray1985}. It is formally equivalent to the Wald-Zoupas charge in
asymptotically-flat general relativity, yet little has been said about the
reason for this connection. The form of the charge arose through careful flux
considerations given previous unsuccessful attempts and the BMS algebra appears
only via a component-wise correspondence between the symmetry algebra and
solutions to the twistor equation.

This article formalizes this correspondence between twistors and the
BMS algebra by first considering a twistor space on a \emph{section}
of null infinity for which no gravitational radiation is arriving.  We
begin with a review of the previous use of twistors in charge
construction in both flat and curved spacetime
(§\ref{sec:spinlowering}). We show that most previous applications of
twistors to charge structure at \(\scri^+\) failed to account for a
certain gauge ambiguity in the twistor which stands in for an element
of the symmetry algebra (§\ref{sec:quiescent}). With a proper
formalism in place, we reinterpret the Dray-Streubel charge procedure
as a more general procedure for regulating the gauge freedom in the
twistor which stands in for an element of the underlying symmetry
algebra so that the resulting charge is adjoint not just to some
twistor space but to the symmetry algebra proper (§\ref{sec:ds}). We
conclude with a discussion of the status of the twistorial description
of the radiative phase space and make an argument for its value as an
alternative description of the radiative observables described by the
shear structure or the symplectic formalism at \(\scri^+\) (§\ref{sec:discussion}).

\subsection{Notation and conventions} \label{sec:notation}
Following Penrose's procedure of conformal compactification
\cite{penrose1965zero}, we consider a spacetime up to a conformal rescaling
\begin{equation}
\label{eq:confmap}
\hat{g}_{\mu\nu} = \Omega^2 g_{\mu\nu}.
\end{equation}
A spacetime may be called asymptotically-flat if it admits a conformal boundary
defined as the level set
\begin{equation}
\label{eq:scridefn}
\scri = \{x \in \mathcal{M}: \Omega(x) = 0\}
\end{equation}
where \(\mathcal{M}\) is the rescaled manifold. It may be shown that \(\scri\) is a null hypersurface with null normal
$N_a = -\nabla_a \Omega$.  Furthermore, we assume $\scri$ is made up of two connected
components with topology \(\mathbb{R} \times S^2 \) called future and past null
infinity. We will work primarily with future null infinity \(\scri^{+}\), but
it is straightforward to demonstrate that all results hold for past null
infinity $\scri^{-}$.

\subsubsection{Tetrad on \texorpdfstring{\(\scri^{+}\)}{I+}}\label{sec:tetrad}
We define a null tetrad on \(\scri\) via a spin frame. Choose a spinor $\iota^A$
such that its flagpole \(n^a \equiv \iota^A \bar{\iota}^{A'}\) is a
representative from the one parameter set of vectors which kill the degenerate
metric on \(\scri\),~i.e.,
\begin{equation}
\label{eq:24}
n_a = -A \nabla_a \Omega
\end{equation}
for some $A$. Note that we
use the abstract index notation of Penrose and so the spinor and tensor indices
need not be thought of as taking on values and instead denote virtual copies of
a given object~\cite{pr1}.

Next, choose a spinor \(o^A\), so that
\begin{equation}
\label{eq:20}
o_A\iota^A = 1
\end{equation}
and its flagpole
\(l^a = o^A \bar{o}^{A'}\), points off \(\scri\). The complex null vector
\(m^a = o^A \bar{\iota}^{A'}\) and its complex conjugate are taken to span
\emph{cuts} of null infinity.

We will regulate the scaling and rotation freedom in the null tetrad using the
GHP formalism \cite{ASpaceTimeCaGeroch1973} on top of the
Newman-Penrose formalism \cite{AnApproachToNewman1962}. In the literature, two special
gauges are employed. The cylinder gauge corresponds to the vanishing of
\(\rho'\) so that each cut shares a unit sphere metric. The Bondi gauge
corresponds to this condition plus the vanishing of \(\tau\) so that the null
lapse is constant across a cut. We will employ a further modified GHP formalism
that is covariant with respect to these gauge transformations.

\subsubsection{Conformal GHP formalism}
The conformal GHP (cGHP) formalism acts on quantities which transform
covariantly with respect to conformal transformations of the metric
\cite{pr2,anewlook}. In that way, it is useful for dealing with quantities on the
\emph{universal radiative structure} at \(\scri^+\) \cite{geroch}.

We say that a scalar \(\eta\) has conformal weight \(w\) if under the conformal
map \(\Omega \to \omega\Omega\) it scales according to
\begin{equation}
\label{eq:13}
\eta \to \omega^w\eta.
\end{equation}
A quantity which is also spin- and boost-weighted will have a weight
denoted \([w;p,q]\). There is the freedom to choose how the null tetrad
scales under such a transformation. We choose
\begin{align}
\label{eq:22}
l^a &\to \omega^{-2} l^a,& m^a &\to \omega^{-1} m^a, & n^a &\to n^a.
\end{align}
As a consequence, the scalar \(A\) has weight \([-1;1,1]\).

One notices that the action of GHP weighted derivatives does not
transform covariantly under \eqref{eq:13} and so we introduce new
conformally-covariant derivatives,
\begin{align}
\label{eq:cghpderivs}
  \thorn'_c \eta &= \thorn'\eta + (w+p+q)\rho' \eta, \\
  \eth_c\eta &= \eth \eta + (w+q)\tau \eta, \\
  \bar{\eth}_c\eta &= \eth' \eta + (w+p)\bar{\tau} \eta,
\end{align}
with corresponding weights
\begin{align}
\label{eq:cghpderivweights}
\thorn'_c \eta &: [w;p-1,q-1], \\ \eth_c\eta &: [w-1;p+1,q-1], \\ \bar{\eth}_c \eta &: [w-1; p-1,q+1].
\end{align}
These derivatives do not commute in general. Their non-commutation is given by,
\begin{align}
\label{eq:ghpcom}
  [\thorn_c',\eth_c]\eta &= (w+q)\mathcal{P} \eta, \\
  [\thorn_c',\bar{\eth}_c]\eta &= (w+p)\bar{\mathcal{P}} \eta, \\
  [\eth, \eth']\eta &= -(p-q) \mathcal{Q} \eta,
\end{align}
where $\mathcal{P} = \thorn' \tau - \eth \rho'$ and
$\mathcal{Q} = \Phi_{11} + \Lambda - \rho \rho' -\eth' \tau$ is related to the the Gauß curvature of the cut.

The components of the rescaled Weyl tensor are properly weighted and their
propagation equations along the generators of \(\scri^+\) are
\begin{equation}
\label{eq:16}
\thorn_c'\psi_k - \eth_c\psi_{k+1} = (3-k)\sigma \psi_{k+2}, \quad k = 0:3.
\end{equation}
The case $k=2$ will be important in later sections.

In a Bondi gauge where cuts are taken to be metric spheres, the Bondi news is
the derivative of \(\sigma\) along the generators, \(\dot{\sigma} = -\bar{N}\),
but the quantity which appears in the Bondi energy in this less restricted gauge is
\(\mathcal{N} = N + \bar{R}\), where \(R\) is the unique solution to
\begin{equation}
\label{eq:cocurvature}
\bar{\eth_{c}} R + \eth_c \mathcal{Q} = 0
\end{equation}
\cite{anewlook,ds}.

One can show that \(\mathcal{N}\) satisfies
\begin{equation}
\label{eq:2}
\thorn_c' \mathcal{N} = A \psi_4, \quad \eth_c \mathcal{N} = A \psi_3
\end{equation}
so that \(\mathcal{N} = N + \bar{R}\) should be taken as the definition of the
Bondi news in this gauge \cite{geroch,ds,anewlook}.

\subsubsection{BMS algebra}
We can write down concrete expressions for elements of the BMS algebra by
considering the Lie dragging of Geroch's universal structure tensor along the
integral curves of a vector field
$X^a = \eta n^a + \bar{\xi} m^a + \xi \bar{m}^a$ on \(\scri\):
\begin{equation}
\label{eq:6}
\mathcal{L}_X {\Gamma^{ab}}_{cd} = 0.
\end{equation}
By taking \(X^a\) to be invariant with respect to conformal, spin, and boost
transformations, the weights corresponding to its components are,
\(\eta : [0;1,1]\) and \(\xi : [1;1,-1]\).
One finds that elements of the BMS algebra are those vector fields $X^a$ such
that
\begin{align}
\label{eq:bmsconstraints}
\eth_c\xi &= 0, & \thorn'_c\xi &= 0, & \thorn'_c \eta &= \frac12 \left( \bar{\eth}_c \xi + \eth_c\bar{\xi} \right).
\end{align}
The value of \(\eta\) on some initial cut is completely unconstrained
so that the BMS algebra is infinite-dimensional. This presentation of
the BMS algebra is that of Frauendiener and Stevens \cite{anewlook}.

It is known that the Bondi-Sachs energy-momentum offers an unambiguous
supertranslation-free definition of energy-momentum on a cut of \(\scri^{+}\) \cite{GravitationalWBondi1973}. The
saving grace which allows for this construction is the existence of an ideal of
translations within the supertranslations \cite{AsymptoticSymmSachs1962}.

In cGHP notation, one finds that the translations are solutions to
\eqref{eq:bmsconstraints} with the further condition,
\begin{equation}
\label{eq:translations}
\eth_c^2 V = RV
\end{equation}
where \(\eta = AV\). It makes sense to talk of a \emph{supertranslation-free}
translation, but the semi-direct product structure of the BMS group means that
there is no obvious way to write down a \emph{supertranslation-free} Lorentz
sub-algebra of the BMS algebra. We must make a choice of supertranslation.

\section{Spin-lowering and twistorial charges}\label{sec:spinlowering}
First, we review the conventional use of twistors in charge construction.
In flat spacetime, twistors provide a very natural description of energy,
momentum, and angular momentum. The charge structure of a system in Minkowski
space is entirely encoded in a symmetric \emph{kinematic} twistor
\(A_{\alpha\beta}\) \cite{pr2}.

One may access the individual charges by contracting with a necessarily
symmetric dual twistor \(S^{\alpha\beta}\). That is,
\begin{equation}
\label{eq:twistorcharge}
A(S^{\alpha\beta}) = A_{\alpha\beta}S^{\alpha\beta},
\end{equation}
so that the choices for
\(S^{\alpha\beta}\) will span the charge structure in the same way that the set
of Killing vectors in flat spacetime spans the charge structure.

Penrose's construction of a quasi-local twistorial charge makes use of the
spin-lowering property of solutions to the twistor equation~\cite{Penrose1982wp}. If we take a spinor solution to the
valence-\(\tvect{2}{0}\) \emph{symmetric} twistor equation,
\begin{equation}
\label{eq:20twistoreqn}
\nabla_{A'}^{(A}\gamma^{BC)} = 0,
\end{equation}
then contracting with a spin-\(s\) zero rest-mass field yields a zero rest-mass
field of spin \(s-1\).

If we contract a spin-\(2\) \emph{graviton} field \(\psi_{ABCD}\), the result is a
spin-\(1\) \emph{Maxwell} field,
\begin{equation}
\label{eq:26}
\phi_{AB} := \psi_{ABCD}\gamma^{CD},
\end{equation}
for which there is a well-motivated charge
construction: choose a relevant 2-surface, contract the area 2-form with the
Maxwell field and integrate over the surface. The independent solutions to
\eqref{eq:20twistoreqn} span the discrepancy between the 1 complex charge in
electromagnetism and the 10 charges we expect in analogy with the linearized gravitational theory.

Furthermore, this procedure is conformally-invariant and so we may construct
such a charge in conformally-flat spacetimes if we take the Weyl spinor as a
spin-\(2\) zero rest-mass field with field equations given by the non-linear
Bianchi identities.

\subsection{Conventional use of twistors on \texorpdfstring{\(\scri^+\)}{I+}}
The work of Dray and Streubel applies this construction to \(\scri^+\)
\cite{ds}. Therefore a prescription for a twistor space on \(\scri^+\)
is needed. A key observation is that when one applies the twistor
equation to \(\scri^+\) only those components of the twistor equation
which are intrinsic to a \emph{cut} of \(\scri^{+}\) are integrable.

Let us consider the valence-\(\tvect{1}{0}\) twistor equation,
\begin{equation}
\label{eq:7}
\nabla_{A'}^{(A}\omega^{B)} = 0,
\end{equation}
and apply it to \(\scri^+\) taking components with respect to the adapted
spin-frame of Section \ref{sec:tetrad}. In terms of cGHP operators,
\begin{equation}
\begin{aligned}
  \bar{\eth}_c \omega^0 &= 0, & \thorn_c'\omega^0 &= 0, \\
  \eth_c \omega^1 &= \sigma \omega^0, & \thorn_c' \omega^1 &= \eth_c \omega^0,
\end{aligned}
\label{eq:10twistor_comp}
\end{equation}
for \(\omega^A = \omega^0 o^A + \omega^1\iota^{A}\), where the
components carry weights \(\omega^0:[1;-1,0]\) and \(\omega^1:[0;1,0]\) if we
choose \(\omega^A\) to have zero weight.

The constraint equation for \(\omega^1\) on a cut does not commute with its
evolution equation along the generators of \(\scri^+\) unless a stringent
integrability constraint is met.

If we would like to make use of the spin-lowering property of solutions to the
twistor equation then we need integrability only on a 2-surface of interest.
Therefore, most authors have taken only those components of the twistor
equation intrinsic to a cut of null infinity, avoiding the
integrability condition \cite{Penrose1982wp,ds,2111.00478v1}.

It is here that one finds the well-known correspondence between twistors and the
BMS algebra of symmetries.

\subsection{Twistor components and the BMS algebra}

According to the procedure of Dray and Streubel, one can apply Penrose's
quasi-local charge definition to cuts of \(\scri^+\) by defining twistors on
these cuts as described above. Then the valence-\(\tvect{2}{0}\) twistor
equation lowers the spin of zero rest-mass fields on this cut by $1$. The
Dray-Streubel charge may claim the title of BMS charge because one finds that
the components of this equation on a cut of \(\scri^+\) match those of an
element of the BMS algebra.

If we define the components of a solution to \eqref{eq:20twistoreqn} on a cut by,
\begin{equation}
\label{eq:10}
\gamma^{AB} = U o^A o^B + V o^{(A}\iota^{B)},
\end{equation}
then the single component of \eqref{eq:20twistoreqn} on a cut is,
\begin{equation}
\label{eq:20twistorcompcut}
  \bar{\eth}_c U= 0.
\end{equation}
Because \(U\) and \(V\) are propagated uniquely along the generators of $\scri^+$
once their values are specified on a cut, the identification,
\begin{align}
\label{eq:3}
\xi &\equiv \bar{U}, & \eta &\equiv V
\end{align}
is made between components of \(\gamma^{AB}\) and components of an element of
the BMS algebra \(X^{a}\) which satisfy \eqref{eq:bmsconstraints}.

One can see that the component $V$ is unconstrained by \eqref{eq:20twistoreqn}
and \(\eta\) is similarly unconstrained since it encodes the
\emph{supertranslation} degrees of freedom of the BMS algebra.

This is the identification made by Dray and Streubel and made explicit in a
recent work \cite{2111.00478v1}. Therefore, solutions to the twistor equation on
a cut are often labeled \emph{BMS twistors}.

This would appear to be an odd
identification to make. Although it is true that both sets of equations span the
same degrees of freedom, they describe two very different objects and the
identification only holds at the level of components. An element of the BMS
algebra is represented by a \emph{vector} \(X^a\) while a solution to
\eqref{eq:20twistoreqn} is a valence-2 spinor. Another objection may be that the
restriction to a single cut is not a well-motivated procedure and is equivalent
to a self-evidently non-covariant selection of preferred components of
\eqref{eq:20twistoreqn}.

Despite these points, the Dray-Streubel charge, which is constructed upon this
identification, exactly aligns with the charge resulting from the Wald-Zoupas
procedure due to entirely different considerations \cite{waldzoupas}. Therefore,
it would be advantageous to provide a physical interpretation for this
identification between the BMS algebra and twistors, leading to a physical
connection between the twistorial analysis and the symplectic procedure of Wald
and Zoupas.

\section{Flat-space twistors and \emph{quiescent} twistors}\label{sec:quiescent}
The physical interpretation we seek is to be found in an isomorphism between
twistors in flat-space and twistors on 3-dimensional \emph{sections} of
\(\scri^+\) for which the Bondi news vanishes, i.e.,~sections for which no
gravitational radiation is arriving at $\scri^{+}$.

The vanishing of the Bondi news for any finite period is unphysical
for a generic dynamical spacetime. It is for this reason that we must
take the infinite-dimensional enhancement from Poincaré to BMS
seriously. Gravitational radiation induces an angle-dependent
\emph{shunting} or \emph{supertranslation} of the frame in which
corresponding observables may be measured. Therefore, it is useful to
consider the stationary case when thinking about constructing a
coherent frame, however we should be clear that this condition is
entirely unphysical and is used only as a way to probe the radiative
phase space.

Let us consider the twistor equation \eqref{eq:7} on such a section. We
have the four equations \eqref{eq:10twistor_comp} intrinsic to \(\scri^+\).
Implicitly, there is also a condition due to the
non-commutativity of the constraint and evolution equations for $\omega^1$. We
can construct it explicitly by attempting to propagate the constraint along the
generators. One finds,
\begin{equation}
\label{eq:23}
\eth_c^2\omega^0 + \bar{N} \omega^0 = 0.
\end{equation}
To see how this condition compares with the explicit constraint
\(\bar{\eth}_c \omega^0 = 0\) we can act with $\eth_c^2$ and commute derivatives
to find,
\begin{equation}
\label{eq:11}
\eth_c^2 \omega^0 - R \omega^{0} = 0.
\end{equation}
where $R$ is the \emph{co-curvature} \eqref{eq:cocurvature} related to the Gaussian curvature of each
cut. Therefore, for the integrability condition to be satisfied, one must have,
\begin{equation}
\label{eq:12}
\mathcal{N} = N + \bar{R} = 0.
\end{equation}
That is, the \emph{Bondi news} as it is defined in this gauge must uniformly
vanish.

This is a well known result and is often given as motivation to consider the
twistor equation only on individual cuts \cite{Penrose1982wp,ds,2111.00478v1}. However, if we do take the
Bondi news to vanish, at least on some portion of null infinity, we are still
left with the constraint \eqref{eq:11}. Then, the twistor equation \eqref{eq:7} on a
radiation-free section of null infinity satisfies,
\begin{equation}
\begin{aligned}
    \eth_c^{2} \omega^0 - R \omega^0 &= 0,& \bar{\eth}_c \omega^0 &= 0,  & \thorn_c'\omega^0 &= 0, \\
    \eth_c \omega^1 - \sigma \omega^0 &= 0, && & \thorn_c' \omega^1 &= \eth_c \omega^0.
\end{aligned}
\label{eq:10twistor_comp_exp}
\end{equation}
By considering expansions of \(\omega^0\) and \(\omega^1\) in spin-weighted
spherical harmonics \cite{NoteOnTheBonNewman1966,EdthADifferenEastwo1982} one can deduce that the solution space of
\eqref{eq:10twistor_comp_exp} is 4-complex-dimensional, matching the degrees of
freedom of the twistor equation \eqref{eq:7} in flat spacetime.

Formally, we may identify the twistor space on a radiation-free
\emph{section} of $\scri^+$ with the twistor space on the
\emph{Minkowski space of origins} corresponding to this section
\cite{211000140v1}. The same cannot be said for a twistor space on a
cut of null infinity since there is no such unique corresponding
Minkowski space of origins. This is a natural consequence of the fact
that $\scri^+$ is a conformal structure shared by \emph{all}
asymptotically-flat spacetimes, and therefore there is no covariant
way to reconstruct the spacetime points in the bulk of any individual
asymptotically-flat spacetime from the conformal structure on
$\scri^+$.

The condition of stationarity (at least for some finite retarded time) picks out
a specific class of spacetimes and admits a shared Minkowski space of
origins between them. This allows for an isomorphism between a twistor
space on a stationary section and flat-space twistors.

We are particularly interested in valence-\(\tvect{2}{0}\) twistors due
to their utility in charge construction as spin-lowering operators, so let us
consider the solution space on a stationary section of $\scri^+$.

The same analysis applies to the valence-\(\tvect{2}{0}\) twistor equation
\eqref{eq:20twistoreqn} as for the valence-\(\tvect{1}{0}\) case. If the
components with respect to the spin-frame on $\scri^+$ are labeled,
\begin{equation}
\label{eq:4}
\gamma^{AB} = U\ o^Ao^B + V\ o^{(A}\iota^{B)} + W\ \iota^A\iota^{B},
\end{equation}
then the 5 components of \eqref{eq:20twistoreqn} with respect to the spin-frame
are,
\begin{equation}
  \begin{aligned}
    \bar{\eth}_c U&= 0, & \thorn_c'U &= 0, \\
                  &    & \thorn_c'V &= \eth_cU, \\
    \eth_cW &= V\sigma, & \thorn_c'W &= \eth_cV -2\sigma U.
  \end{aligned}
\label{eq:20twistor_comp}
\end{equation}
where the components of \(\gamma^{AB}\) have weights, \(U: [2;-2,0]\),
\(V:[1;0,0]\), and \(W:[0;2,0]\) if we take \(\gamma^{AB}\) to have zero weight.

Again, there is an implicit constraint due to integrability along the generators
of $\scri^+$, this time due to the non-commutativity of the constraint and
evolution equations for the component $W$. Again we find that the Bondi news
$\mathcal{N}$ must vanish and there is the implicit constraint,
\begin{equation}
\label{eq:14}
\eth_c^2 V - 2U \eth_c \sigma - 3 \sigma \eth_c U - RV = 0.
\end{equation}

This constraint bears an informal resemblance with the condition which picks out
the four-dimensional ideal of translations from the supertranslations
\eqref{eq:translations}.

The constraint \eqref{eq:14} may be regarded as a deformation of this condition
which we will discuss in more detail in Section \ref{sec:quiescentflat},
but for now it suffices to point out that the solution space degrees
of freedom of \eqref{eq:20twistor_comp} and \eqref{eq:14} match those
of the valence-\(\tvect{2}{0}\) twistor equation in flat
spacetime. This can be confirmed formally via the Atiyah-Singer index
theorem or informally using an expansion in spin-weighted spherical
harmonics.

What we have is a true \emph{isomorphism} between twistors in Minkowski
spacetime and twistors on \emph{radiation-free} sections of
\(\scri^+\) and not just a selection of components of the twistor
equation. This will be the key observation towards formalizing the
connection between twistors and asymptotic symmetries because a
certain class of twistors in Minkowski spacetime encode the Killing
vectors \emph{of Minkowski spacetime}. We will see that this
connection may be translated to the isomorphic twistor spaces on
\(\scri^+\).

\subsection{Twistors and symmetries}\label{sec:twistsym}
We typically regard a charge as adjoint to some group of symmetries. In the
quasi-local charge construction of Penrose, no such symmetry group appears. The
degrees of freedom are spanned by twistorial degrees of freedom.

In fact, there
is a well-defined connection between twistors and symmetries in flat spacetime.
The link is not direct but passes through a gauge freedom in the twistor which
represents a given element of the Poincaré algebra \cite{pr2}. We will see that this link
can be translated to the isomorphic twistor spaces on stationary sections of
$\scri^+$.

The natural twistorial charge structure in flat spacetime
\eqref{eq:twistorcharge} is related to the algebra of Killing vectors of Minkowski
space. If the spinor components of \(S^{\alpha\beta}\) are denoted
\begin{equation}
\label{eq:1}
S^{\alpha\beta} = \begin{bmatrix} \sigma^{AB} & {\rho^A}_{B'} \\ {\tau_{A'}}^B & \kappa_{A'B'} \end{bmatrix},
\end{equation}
then the component \(\rho^{AB'}\) is a \emph{complex} Killing vector which has a real
part that satisfies Killing's equation. \(\rho^{AB'}\) is related to the primary
component of \(S^{\alpha\beta}\) by
\begin{equation}
\label{eq:twisttokill}
\nabla_{CC'} \gamma^{AB} = -2i {\epsilon_C}^{(A}{\rho^{B)}}_{C'}.
\end{equation}
It is quite crucial that this twistor does not determine a Killing vector
directly but through a \emph{complex} Killing vector. Since more than one
twistor may define a \(\rho^{AB'}\) whose real part is a given Killing vector,
the space of symmetric twistors $S^{\alpha\beta}$ is strictly larger than the
space of Killing vectors. Explicitly, a given Killing vector may be represented
by a twistor $S^{\alpha\beta}$ not uniquely but up to a transformation,
\begin{equation}
\label{eq:96}
S^{\alpha\beta} \to S^{\alpha\beta} + 2i {G^{(\alpha}}_{\gamma}I^{\beta)\gamma},
\end{equation}
for an arbitrary Hermitian twistor
${G^{\alpha}}_{\beta} = \tensor{\bar{G}}{_{\beta}^{\alpha}}$ \cite{pr2}.

This relationship between twistors and symmetries may be carried over to the
isomorphic twistor space on stationary sections of $\scri^+$ meaning that the
solution space of \eqref{eq:20twistor_comp} and \eqref{eq:14} spans the degrees of freedom of the
Poincaré algebra \emph{up to a gauge transformation \eqref{eq:96}}.

If we denote components of \(\rho^{AB'}\) with respect to the spin-frame by
\begin{equation}
\label{eq:5}
\rho^a = \eta n^a + \bar{\xi} m^a + \xi \bar{m}^a,
\end{equation}
then at the level of components the link between twistors and symmetries takes
the form of an identification,
\begin{align}
\label{eq:8}
\bar{\xi} &= AU, & \eta &= AV,
\end{align}
where $U$ and $V$ are the components of $\gamma^{AB}$.

The ambiguity \eqref{eq:96} is reflected in the condition that only the
\emph{real} part of $\rho^a$ corresponds to the symmetry algebra degrees of
freedom. At the level of components, $\text{Im}(V)$ should not appear in any
twistorial quantity which describes a property related to a symmetry.

Penrose's quasi-local charge prescription defines a set of charges adjoint to
the space of valence-$\tvect{2}{0}$ twistors on a 2-dimensional cut by
exploiting the spin-lowering property of twistors. If such a charge purports to
be adjoint to this underlying symmetry group also, we must ensure that it is
invariant under the actions of the gauge transformations \eqref{eq:96}. We will
see that the Dray-Streubel charge may be reinterpreted as a procedure for
regulating this gauge freedom to define a charge adjoint to the symmetry group
proper.

\subsection{Quiescent twistors and flat
  spacetime}\label{sec:quiescentflat}
Let us say more about translating the indirect connection between twistors and symmetries described in
Section \ref{sec:twistsym} to twistor spaces defined on stationary or
\emph{quiescent} sections of \(\scri^+\).

To proceed, consider a spacetime for which the Bondi news vanishes
initially for some finite retarded time before a period of radiation
arrives at \(\scri^{+}\). In this initial quiescent period
\(\mathcal{S}_{0}\), we can define a quiescent twistor space
isomorphic to flat-space twistors. In \(\mathcal{S}_0\), we have
\(\sigma = 0\) and the constraint \eqref{eq:14} becomes
\begin{equation}
\label{eq:15}
\eth_c^2 V  - RV = 0.
\end{equation}
This is exactly the condition which picks out the ideal of translations from the
supertranslations to distinguish the Poincaré algebra from the BMS
algebra. In fact, under the identification \eqref{eq:8}, this
condition becomes exactly the condition \eqref{eq:translations}.

Now, if a burst of gravitational radiation were to pass, displacing \(\sigma\)
and delimiting the start of a new radiation-free region, it would disrupt the
integrability of the twistor equation, but if we impose \eqref{eq:20twistoreqn}
on the new radiation-free region we have the constraint \eqref{eq:14} with
\(\sigma \neq 0\) in general.

If we consider a series of these \emph{quiescent} regions
\(\mathcal{S}_0,\ \mathcal{S}_1,\ \mathcal{S}_2,\ \ldots\) interspersed
with bursts of radiation, each one will have an independent twistor
space \(\mathcal{T}_0,\ \mathcal{T}_1,\ \mathcal{T}_2,\ \ldots\) which is
each exactly isomorphic to the twistor space of flat
spacetime. However, there is no canonical map between each of
these \(\mathcal{T}_n\) so that we \emph{do not} have a shared twistor
space which may be used to define a supertranslation-free angular
momentum or something similar on either side of a burst of
radiation. The \(\mathcal{T}_n\) are \emph{relatively
  supertranslated}.

The reason one should consider these spaces despite the unphysical
nature of the absence of gravitational radiation is that through the
isomorphism between \(\mathcal{T}_n\) and flat spacetime twistors, we
can pull the connection between flat spacetime twistors and Killing
vectors \emph{of Minkowski spacetime} to null infinity so that each
\(\mathcal{T}_n\) carries its own set of Killing vectors \emph{of
  Minkowski spacetime} subject to the gauge discrepancy
\eqref{eq:96}.

To summarize, each quiescent region \(\mathcal{S}_{n}\) has a
corresponding quiescent twistor space \(\mathcal{T}_n\) to which there
is a corresponding \emph{copy} of the Poincaré algebra \(\mathcal{P}_n\) through the
connection described in Section \ref{sec:quiescent}. However, because
there is no canonical map between each of the
\(\mathcal{T}_n\), there can be no canonical map between the
Poincaré subalgebras \(\mathcal{P}_n\).

In general, the \(\mathcal{P}_n\) are relatively supertranslated
Poincaré subalgebras of the BMS algebra. Each subalgebra shares only
the \emph{translations} with the others which can be seen by
considering the pure supertranslations which meet the constraint
\eqref{eq:14}. This situation has been described by Penrose and
Rindler but here we arrive at it twistorially \cite{pr2}.

In each case the \emph{local} integrability of the twistor equation in
the absence of radiation gives rise to an isomorphism between the
\(\mathcal{T}_n\) and flat-space twistors which in turn gives access
to a \emph{global} property, a Killing vector. Symmetry-adjoint
charges are part of the ontology of a theory constructed on top of
the global structure of Minkowski spacetime. The above interplay between the local and global
gives us some insight into why we are able to define a charge in
curved spacetime.

\subsection{BMS twistors}
The above analysis demonstrates the connection between a twistor space on a
section of $\scri^+$ for which there is no arriving radiation and the Poincaré
algebra. What is the connection to the Dray-Streubel charge which is adjoint to
the infinite-dimensional BMS algebra of symmetries?

We may recover the BMS algebra from this construction by restricting
to a cut and observing the corresponding effect on an element of the
Poincaré algebra under the map \eqref{eq:twisttokill}. This removes
the integrability condition on the twistor equation which imposed an
absence of radiation and picked out the translations. We are left with
the unconstrained supertranslations.

Twistors on a cut of $\scri^+$ have been called \emph{BMS twistors}
since the solution space degrees of freedom of the twistor equation on
a cut match those degrees of freedom of the BMS algebra
\cite{2111.00478v1}. There is now a convincing formalization of the
connection between twistors and the asymptotic symmetry algebra in the
contraction of a twistor space on all of $\scri^+$ to a cut and the
effect of this contraction on the spinor component of the twistor
which represents a \emph{complex} Killing vector \emph{of Minkowski
  space}.

Importantly, the same gauge freedom \eqref{eq:96} holds for a \emph{BMS twistor}
which purports to stand in for an element of the BMS algebra. We will now see
that the Dray-Streubel charge may be reinterpreted as exactly the necessary
procedure to regulate this gauge freedom in a twistorial charge.

\section{The Dray-Streubel charge prescription}\label{sec:ds}
We now demonstrate that the charge of Dray and Streubel \cite{ds} may be thought
of as a more general procedure to define a charge which is adjoint to a space of
twistors at null infinity but which regulates the gauge freedom in the choice of
twistor which stands in for an element of the symmetry algebra. This regulation is done by the
introduction of a vanishing term with a non-vanishing flux and the resulting charge
may be identified as adjoint to the symmetry algebra proper since these gauge
degrees of freedom are constrained.

The details of the procedure are as follows. One may define a natural
gravitational charge on a 2-surface by first lowering the spin-$2$ Weyl spinor
to a spin-$1$ Maxwell field $\phi_{AB}$ using a symmetric
valence-\(\tvect{2}{0}\) twistor with primary spinor component denoted
$\gamma^{AB}$. Then,
\begin{equation}
\label{eq:17}
\phi_{AB} := \Psi_{ABCD}\gamma^{CD}
\end{equation}
is a Maxwell field since it satisfies the zero rest-mass field equations
\cite{Penrose1982wp}.

There is a natural charge definition for Maxwell fields given by contraction
with the area 2-form of the surface and integrating. This is Penrose's
quasi-local charge prescription. Since the degrees of freedom will be entirely
contained in the choice of twistor, we say that the resulting charge is a charge
adjoint to the space of twistors $\gamma^{AB}$. Explicitly, we have
\begin{equation}
\label{eq:18}
A(\gamma^{AB}) = \oint_{\Sigma} \Psi_{ABCD} \gamma^{CD} o^{A} \iota^{B}\ dS,
\end{equation}
where $\Sigma$ is the 2-surface on which the charge is evaluated and
\(o^{A} \iota^{B}\ dS\) is the natural area 2-form on $\Sigma$.

On \(\scri^{+}\) with respect to the natural spin-frame, \eqref{eq:18} takes the form,
\begin{align*}
\label{eq:19}
A(\gamma^{AB}) &= \oint_{\Sigma} U\psi_1 + V\psi_2 + W\psi_3\ dS \\ &= \oint_{\Sigma} U\psi_1 + V\left( \psi_2 - A^{-1} \sigma \mathcal{N} \right)\ dS \addnumber
\end{align*}
where $U$, $V$, and $W$ are components of $\gamma^{AB}$ with respect to the
spin-frame and we have integrated the third term by parts to arrive at the
second equality.

When the space of twistors is defined on a single cut, this is Penrose's elegant
definition of a twistorial charge. The space of symmetric
valence-\(\tvect{2}{0}\) twistors spans its degrees of freedom. However, \eqref{eq:19}
is not invariant with respect to map \eqref{eq:96} and so it cannot yet be
thought of as a charge adjoint to the corresponding symmetry algebra.

To define a \emph{symmetry-adjoint} charge, we should take the
component of Penrose's twistorial charge which is invariant with
respect to \eqref{eq:96}, the gauge freedom in the twistor which
stands in for an element of the symmetry algebra. At the level of
components this corresponds to the condition that \(\text{Im}(V)\)
does not appear in the charge expression. One can see that taking the
real part of the charge \eqref{eq:19} does not immediately guarantee
that \(\text{Im}(V)\) will not appear since the coefficient of \(V\)
in \eqref{eq:19} is not necessarily real.

The key insight is the existence of the integral
\begin{equation}
\label{eq:21}
\oint_\Sigma V \left[ \bar{\eth}_{c}\sigma - \bar{R} \sigma \right] dS = 0.
\end{equation}
The vanishing of \eqref{eq:21} may be shown by integrating by parts and applying the
constraint on $V$.

We may add any multiple $q$ of \eqref{eq:21} to the charge expression
\eqref{eq:19} so that,
\begin{equation}
\label{eq:chargeflux1}
A(\gamma^{AB}) = \oint_{\Sigma} U\psi_1 + V\left[ \psi_2 - A^{-1}\sigma \mathcal{N} -q \left(\bar{\eth}_{c}\sigma - \bar{R} \sigma \right) \right]\ dS.
\end{equation}
then the \emph{real} part of \(A(\gamma^{AB})\) contains no reference to \(\text{Im}(V)\) so
long as
\begin{equation}
\label{eq:imagcondition}
\text{Im} \left( \psi_2 - A^{-1}\sigma \mathcal{N}  -q \left(\bar{\eth}_{c}\sigma - \bar{R} \sigma \right)\right) = 0.
\end{equation}
From the propagation equations for Ricci components \eqref{eq:16}, this condition
is satisfied if $q=1$.
Then \(\text{Re}(A(\gamma^{AB}))\) with the inclusion of the above flux term is
invariant with respect to \eqref{eq:96} and is therefore adjoint to the symmetry
algebra proper.

Dray and Streubel identify a second vanishing term of non-vanishing flux,
\begin{equation}
\label{eq:chargeflux2}
\oint_\Sigma \left[ U\left(2\sigma\eth_{c}\bar{\sigma} + \eth_{c}(\sigma\bar{\sigma})\right)
+ V \left( \eth_c^{2}\bar{\sigma}-R\bar{\sigma}\right) \right] dS = 0,
\end{equation}
which has the ability to shift flux between the \emph{supermomentum} and angular
momentum components of the charge, naively identified. One may instead add a
linear combination of \eqref{eq:21} and \eqref{eq:chargeflux2} to the
charge expression constrained by $p+q = 1$. The identification by Dray and
Streubel of this ability to shift flux is what resolved the nefarious
\emph{factor of two} discrepancy suffered by earlier charge prescriptions.

Including this second flux term, one is left with the Dray-Streubel
charge when the twistor space is taken to be twistors defined on a cut
of null infinity.  However, for Dray and Streubel, the motivation for
including the flux term \eqref{eq:21} in their charge
expression was that it gave a real supermomentum flux for real
supertranslations. Their final charge was constructed by splitting the
angular momentum and supermomentum components of the twistorial charge
and specifying
\begin{equation}
\label{eq:9}
Q_{\text{DS}} = Q_{V} + Q_{U} + \bar{Q}_{U},
\end{equation}
where subscripts denote the term in the charge with the corresponding
coefficient. This form was chosen to align with previous attempts at charge constructions.

The above procedure shows that in fact the flux term \eqref{eq:21} plays the
crucial role in the Dray-Streubel construction since it is what allows for
the charge to be adjoint to the symmetry algebra. Furthermore, we are led to
consider the real part of the twistorial charge in service of this same
condition. The formal connection between twistors and the asymptotic
symmetry algebra resolves the essential ad hoc elements of the
Dray-Streubel procedure.

When the twistor space is the space of twistors on a \emph{cut} of $\scri^+$ then
\(\text{Re}(A(\gamma^{AB}))\) is the Dray-Streubel BMS charge, but we can now
consider other twistor spaces on $\scri^+$, in particular, the
\emph{quiescent} twistors defined on radiation-free sections of \(\scri^+\).
It is clear that the corresponding Dray-Streubel charge will be adjoint to the
Poincaré algebra by the discussion of Section~\ref{sec:twistsym}.

With a formalization of the connection between twistors and the
asymptotic symmetry algebra, the task of defining a BMS charge follows
naturally from the known relationship between twistors and symmetries
in flat spacetime. It is surprising that the charge which one arrives
at by this simple procedure matches the Wald-Zoupas charge derived
from the explicit symplectic structure of the radiative phase space \cite{waldzoupas}.
Perhaps the twistor formalism can function more generally as a tool
for working with the symplectic structure at \(\scri^+\).

\section{Discussion}\label{sec:discussion}
The construction of \emph{quiescent} twistors requires a finite
region for which there is no gravitational radiation arriving so that
the frame in which the charge is measured may be determined. One might
imagine that for \(u \to -\infty\) and \(u \to \infty\), such regions
may exist, perhaps in a limiting sense, for a certain class of
radiating spacetimes. Likely, this is still too stringent a
constraint to place on physically reasonable spacetimes.

The utility of studying such unphysical spacetimes is that they turn
the problem of \emph{strong} supertranslation ambiguity where
radiation imparts a continuous angle-dependent shunting of the frame
into a number of discrete supertranslations induced by bursts of
radiation between a number of \emph{quiescent} regimes for which no
radiation arrives. From this point of view, it is clearer to see the
the relationship between arriving radiation, shear structure, and the
appearance of the global structure of Minkowski spacetime at
\(\scri^+\).

A charge structure is necessarily a global structure because it relies
on a covariant phase space from which to select values. In flat
spacetime this structure is represented by the existence of a set of
Killing vectors to which one may define a set of adjoint
charges. Twistorial structure is fundamentally connected to the
\emph{global} structure of Minkowski space and so on \(\scri^+\),
\emph{local} twistor integrability carries global properties of
Minkowski space like its set of Killing vectors to the radiative phase
space. Furthermore, the breaking of integrability along \(\scri^+\)
exactly corresponds to the widening of Poincaré to BMS. The
Dray-Streubel charge exploits this interplay between the local and
global.

We would like to suggest that twistors do more than just provide a
useful trick for defining a charge in asymptotically-flat
spacetimes. Since they have the ability to reproduce the BMS charge
expression of the symplectic formalism and their integrability along
the generators of \(\scri^+\) directly mirrors shear structure, they
may provide a framework for analyzing the phase space of radiative
observables on par with these standard descriptions of radiative
data at \(\scri^{+}\). Previously, the use of twistors here was limited by the ad hoc
connections between twistors and the asymptotic symmetry group. With a
proper formalization of this connection, the value of the twistorial
description of the radiative phase space is clear.

Future work will explore the ability for this twistorial description
to reproduce constructions in the symplectic formalism in the vein of the
correspondence between the Wald-Zoupas and Dray-Streubel charges. Perhaps it is
possible to call upon the considerable suite of tools from twistor theory to
answer questions that arise in the symplectic description of the
radiative phase space.

\begin{acknowledgments}
The author would like to thank J\"org Frauendiener for useful correspondence.
\end{acknowledgments}

\bibliography{refs}
\end{document}